%% file: main.tex
\documentclass[runningheads]{llncs}
\usepackage{graphicx}
\usepackage[ruled]{algorithm}
\usepackage{algorithmicx}
\usepackage{algpseudocode}
\usepackage{booktabs}
\usepackage{xcolor}
\usepackage{listings}
\usepackage{amsmath}
\usepackage{indentfirst}
\usepackage{multirow}
\usepackage{amsfonts}
\usepackage{mathrsfs,amsmath}
\usepackage{diagbox}
\usepackage{booktabs}  
\usepackage{threeparttable} 
\usepackage{amssymb}
\usepackage{wasysym}
\usepackage{makecell}
\usepackage{array}
\usepackage[colorlinks,linkcolor=blue]{hyperref}
\usepackage{marvosym}
\begin{document}
\begin{sloppypar}
%
\title{Faster Post-Quantum TLS 1.3 Based on ML-KEM: Implementation and Assessment}
%
%
\author{Jieyu Zheng \and
Haoliang Zhu \and Yifan Dong \and Zhenyu Song \and
Zhenhao Zhang \and Yafang Yang \and Yunlei Zhao$^{\href{mailto:ylzhao@fudan.edu.cn}{\textrm{\Letter}}}$}
\authorrunning{J. Zheng et al.}
%
\institute{School of Computer Science and Technology, Fudan University, Shanghai, China \\
\email{\{jyzheng23,hlzhu22,yfdong22,songzy23,zhenhaozhang23\}@m.fudan.edu.cn, \{18110240046,ylzhao\}@fudan.edu.cn}}
\maketitle              
\begin{abstract}

TLS is extensively utilized for secure data transmission over networks. However, with the advent of quantum computers, the security of TLS based on traditional public-key cryptography is under threat. To counter quantum threats, it is imperative to integrate post-quantum algorithms into TLS. Most PQ-TLS research focuses on integration and evaluation, but few studies address the improvement of PQ-TLS performance by optimizing PQC implementation.\\
 {\setlength{\parindent}{12pt}\indent For the TLS protocol, handshake performance is crucial, and for post-quantum TLS (PQ-TLS) the performance of post-quantum key encapsulation mechanisms (KEMs) directly impacts handshake performance. In this work, we explore the impact of post-quantum KEMs on PQ-TLS performance. We explore how to improve \textsf{ML-KEM} performance using the latest Intel's Advanced Vector Extensions instruction set AVX-512. We detail a spectrum of techniques devised to parallelize polynomial multiplication, modular reduction, and other computationally intensive modules within \textsf{ML-KEM}. Our optimized \textsf{ML-KEM} implementation achieves up to 1.64$\times$ speedup compared to the latest AVX2 implementation. Furthermore, we introduce a novel batch key generation method for \textsf{ML-KEM} that can seamlessly integrate into the TLS protocols. The batch method accelerates the key generation procedure by 3.5$\times$ to 4.9$\times$. We integrate the optimized AVX-512 implementation of \textsf{ML-KEM} into TLS 1.3, and assess handshake performance under both PQ-only and hybrid modes. The assessment demonstrates that our faster ML-KEM implementation results in a higher number of TLS 1.3 handshakes per second under both modes.  Additionally, we revisit two IND-1-CCA KEM constructions discussed in Eurocrypt22 and Asiacrypt23. Besides, we implement them based on  \textsf{ML-KEM} and integrate the one of better performance into TLS 1.3 with benchmarks. }

\end{abstract}
\section{Introduction}

Digital communications are ubiquitous worldwide, with most Internet connections relying on Transport Layer Security (TLS) to secure data transmission. However, the current TLS protocol remains vulnerable to quantum attacks.  TLS employs public-key cryptography algorithms, including Elliptic Curve Diffie-Hellman (ECDH), Elliptic Curve Digital Signature Algorithm (ECDSA), and RSA. However, these algorithms are susceptible to quantum computing threats, as demonstrated by Shor's algorithm \cite{shor1994algorithms}. To address this vulnerability, the National Institute of Standards and Technology (NIST) launched the Post-Quantum Cryptography (PQC) competition in 2016. After three rounds,  NIST announced the selection of the first algorithms to be standardized, including \textsf{Kyber}, \textsf{Dilithium}, \textsf{Falcon}, and \textsf{SPHINCS+} \cite{NIST2022}. In August 2023, NIST designated three standard drafts:  \textsf{ML-KEM} \cite{NISTFIPS203}, \textsf{ML-DSA} \cite{NISTFIPS204}, and \textsf{ML-SLH} \cite{NISTFIPS205}, renamed from  \textsf{Kyber}, \textsf{Dilithium} and \textsf{SPHINCS+}  respectively.

Amidst the continuous progress in NIST's PQC algorithm standardization, significant endeavors have been devoted to the development of PQ-TLS over the past decade. Beginning in 2014, Chang et al. \cite{chang2014postquantum} introduced a post-quantum SSL/TLS library for embedded systems. Subsequently, Bos et al.  \cite{bos2015post} developed PQC ciphersuites for TLS based on the ring learning with errors (R-LWE) problem. In July 2016, Chrome introduced the \textsf{newhope1024} post-quantum option, signaling an initial step towards integrating post-quantum cryptography into mainstream browsers. However, due to patent issues, Chrome removed the \textsf{newhope1024} option in November 2016. Subsequent efforts by industry giants such as Google and Cloudflare focused on evaluating the performance of post-quantum cryptographic candidates within TLS. In 2018, Google and Cloudflare conducted experiments with NIST PQC candidate \textsf{HRSS} and \textsf{X25519} in TLS 1.3 Chrome, while in 2019 they utilized \textsf{ntruhrss701} and \textsf{sntrup761} in their TLS experiments \cite{CECPQ2}.

Recent research on PQ-TLS has predominantly concentrated on three pivotal domains:

\begin{itemize}
    \item \textbf{Integration of PQC into the TLS Protocol:} Exploring methodologies to seamlessly incorporate post-quantum cryptographic mechanisms into the TLS framework \cite{crockett2019prototyping,schwabe2021more,schwabe2020post,paul2021tpm,paul2022mixed,garcia2023quantum,pablos2022design}.

    \item \textbf{Performance Evaluation and Communication Overheads:} Assessing the computational efficiency and communication overheads incurred by post-quantum cryptographic primitives within the TLS ecosystem \cite{henrich2023performance,sosnowski2023performance,bozhko2023performance,tasopoulos2022performance,gonzalez2022kemtls,fan2021impact,DBLP:conf/ndss/SikeridisKD20,paquin2020benchmarking,DBLP:conf/conext/SikeridisKD20,DBLP:journals/access/Abiega-LEglisse20}.

    \item \textbf{Optimized Implementations for Enhanced PQ-TLS Efficiency:} Enhancing the computational efficiency of post-quantum cryptography within TLS through optimized implementations \cite{DBLP:conf/uss/BernsteinBCT22}.
\end{itemize}

Most research on PQ-TLS primarily focuses on exploring how to integrate various PQC cryptographic primitives into TLS and evaluating their performance. However, there is a noticeable scarcity of work improving PQ-TLS performance. Performance stands as a critical factor in TLS applications. As an integral component of PQ-TLS, PQC algorithms directly impact the handshake time of PQ-TLS. Therefore, optimizing the implementation of PQC algorithms and integrating them into TLS may contribute to reducing the handshake time of PQ-TLS.

\paragraph{Motivations.} Recent work presented an accelerated \textsf{Ed25519} and \textsf{X25519} AVX-512 engine tailored for TLS 1.3, offering significant performance improvement \cite{zhang2024eng25519}. Previous work explored optimized PQC implementations using the latest Intel Single Instruction Multiple Data (SIMD) instruction AVX-512 (e.g. \cite{cheng2022highly,lei2023faster,cheng2021batching}). However, the integration of PQC AVX-512 optimized implementations into TLS 1.3 remains unrealized, and there is currently no AVX-512 implementation available for \textsf{ML-KEM}. This gap prompts our focus on optimizing \textsf{ML-KEM} using AVX-512 and seamlessly integrating the optimized \textsf{ML-KEM} implementation into TLS 1.3, and also provides an opportunity to explore the impact of AVX-512 instructions on PQ-TLS handshake protocols. 

As of now, existing research on PQ-TLS migration relies on OQS-OpenSSL, which lacks support for OpenSSL3 and remains outdated. However, the latest OQS provider \cite{oqsprovider23} not only supports OpenSSL3 but also facilitates a clear separation between OpenSSL code and PQC KEM code. Our comprehensive integration of \textsf{ML-KEM} using the OQS provider could provide valuable guidance for researchers seeking to migrate to PQ-TLS.

In addition, recent studies \cite{DBLP:conf/eurocrypt/Huguenin-Dumittan22,jiang2023post,schwabe2020post,schwabe2021more} have affirmed the sufficiency of IND-1-CCA KEMs for TLS 1.3 handshake to be secure. Intriguingly, IND-1-CCA KEMs can be obtained from any OW-CPA/IND-CPA KEMs without re-encryption and de-randomization \cite{DBLP:conf/eurocrypt/Huguenin-Dumittan22,jiang2023post} that are required by  IND-CCA KEMs used in PQ-TLS previously. An idea can be easily deduced that TLS 1.3 handshake might demonstrate improved efficiency by applying such IND-1-CCA KEMs. However, there remains a notable absence of experiments focusing on IND-1-CCA-security KEM TLS 1.3. This gap in research catalyzes us to conduct experiments and evaluations on IND-1-CCA-secure PQ-TLS  handshake protocols.

\paragraph{Contributions.}  In this work, we aim to bridge the gap between PQC engineering implementations and TLS protocol applications. We approach to this task from both an optimization engineering perspective and a TLS system perspective.  We will later open source our code.


\begin{itemize}

\item We present the first optimized implementation of \textsf{ML-KEM} using AVX-512.  As the main bottleneck in \textsf{ML-KEM} lies in polynomial multiplication and hash functions, we achieve 32-way parallel polynomial multiplication and 8-way hash function. Besides, we enhance polynomial rejection and central binomial distribution sampling through the new features of AVX-512 like masked registers and compressive store instructions. Our implementation successfully passes NIST's KAT tests, achieving a 1.64$\times$ speedup compared to the state-of-the-art AVX2 implementation of ML-KEM.

    \item We propose a batch key generation method for \textsf{ML-KEM} to batch 8 independent key pairs. Our batch key generation method achieves a speedup of 3.5$\times$ to 4.9$\times$ compared to key generation without batching. This batch generation approach can also be applied to other key generation processes involving hash function calls.

    \item We revisit two IND-1-CCA KEM constructions discussed in Eurocrypt’22 \cite{DBLP:conf/eurocrypt/Huguenin-Dumittan22}  and Asiacrypt’23 \cite{jiang2023post},  and implement them with the underlying CPA-secure  PKE of \textsf{ML-KEM}.  We then evaluate the performance of IND-1-CCA KEMs, and integrate the better one into TLS 1.3. The benchmark results indicate that IND-1-CCA KEMs improve the performance of the TLS 1.3 handshake compared to IND-CCA KEMs.

    \item We integrate the AVX-512 optimized implementation of \textsf{ML-KEM} into TLS 1.3, assess its impact on TLS 1.3 handshake time, and evaluate the influence of different KEM constructions on TLS handshake efficiency. Our evaluation reveals that an efficient implementation of ML-KEM utilizing AVX-512 can yield a higher number of handshakes per second compared to the latest AVX2 implementation.

\end{itemize}
\section{Preliminaries}
\subsection{Notation}
The notation in this paper is the same as the FIPS 203 draft \cite{NISTFIPS203}. We denote $\mathcal{R}_q$ as the cyclic polynomial ring $\mathbb{Z}_q[x]/(x^n+1)$. We define $r'= r~{\bmod}^{\pm}~\alpha$ (resp. $r' = r \bmod \alpha$) to be the unique element $r'$ in the range $-\left\lfloor{\frac{\alpha}{2}}\right\rfloor<r^{\prime} \leq \left\lfloor{\frac{\alpha}{2}}\right\rfloor$ (resp. $0 \leq r^{\prime}<\alpha$) such that $\alpha |(r - r')$. By default, regular font letters
denote elements in $\mathcal{R}_{q}$, bold lower-case letters are column vectors and bold upper-case letters are matrices.
\subsection{ML-KEM}
\textsf{ML-KEM} is a NIST-standardized lattice-based KEM. Its security is based on the Module Learning With Errors (M-LWE) problem. \textsf{ML-KEM} is derived from Round 3 version of \textsf{Kyber} \cite{avanzi2019crystals}. The polynomial multiplication over $\mathbb{Z}_{3329}[x] /\left(x^{256}+1\right)$  is a fundamental operation in \textsf{ML-KEM}. Utilizing the property $n|(q-1)$, \textsf{ML-KEM} employs an incomplete Number Theoretic Transform (NTT) to accelerate this operation. \textsf{ML-KEM} uses four SHA-3 hash functions: \textsf{SHA3-256}, \textsf{SHA3-512}, \textsf{SHAKE128}, and \textsf{SHAKE256}. For more details, readers can refer to FIPS 203 draft \cite{NISTFIPS203}. \\
\textbf{Number Theoretic Transform.}
NTT is a variant of Fast Fourier Transform (FFT) in finite fields. Its essence is to use the point-value representation of polynomials to perform efficient polynomial multiplication operations. We denote the forward NTT transform as \textsf{NTT}, and the inverse NTT as \textsf{INTT}. The symbol ``$\cdot$" denotes point-wise multiplication. Polynomial multiplication $h(x)=f(x) \times g(x) \in \mathbb{Z}_q[x]/ \left(x^n+1\right)$ can be computed as follows:
$$
h(x)=f(x) \times g(x)=\operatorname{INTT}(\operatorname{NTT}(f) \cdot \operatorname{NTT}(g)).
$$
For cyclic polynomial ring $\mathbb{Z}_q[x] /\left(x^n+1\right)$, complete NTT requires $2n|(q-1)$, which is not satisfied in \textsf{ML-KEM} parameters. Instead, \textsf{ML-KEM} uses a variant of NTT, which deletes the last layer of the NTT. For this type of NTT that follows the ``bottom cropping" method, we call it T-NTT (truncated NTT) for short and let $\beta$ be the number of truncated layers. \textsf{ML-KEM} uses the case of $\beta=1$ for T-NTT. Given $\mathbb{Z}_q [x]/(x^n+1)$, for any integer $\beta>=0$, $q$ is a prime satisfies  
$\frac{n}{2^{\beta-1}} \mid(q-1)$, T-NTT $(f)$ has the general form:
\[\mathbb{Z}_q[x]/\left( {{x^n} + 1} \right) \cong \prod\limits_{i = 0}^{\frac{n}{{{2^\beta }}} - 1} {\mathbb{Z}_q[x]/\left( {{x^{{2^\beta }}} - \omega _{\frac{{2n}}{{{2^\beta }}}}^{2 \cdot b{r_{n/{2^\beta }}}(i) + 1}} \right)} \]
where ${\omega}_{2n/2\beta}$ denotes the $\frac{2n}{{{2^\beta }}}$-th root of unity, $b{r_{n/{2^\beta }}}(i)$ denotes the bit-reversal permutation of $\{0,1, \ldots ,\frac{n}{{{2^\beta }}}-1\}$.


Through the T-NTT transformation, the $n$-dimensional polynomials $f,g$ are separately decomposed into $\frac{n}{2}$ linear polynomials. Then the original multiplication $f \times g$ is transformed into $\frac{n}{2}$ multiplications of linear polynomials.
\begin{figure}[htbp]
\centering
\begin{minipage}[t]{0.48\textwidth}
\centering
\includegraphics[width=6cm]{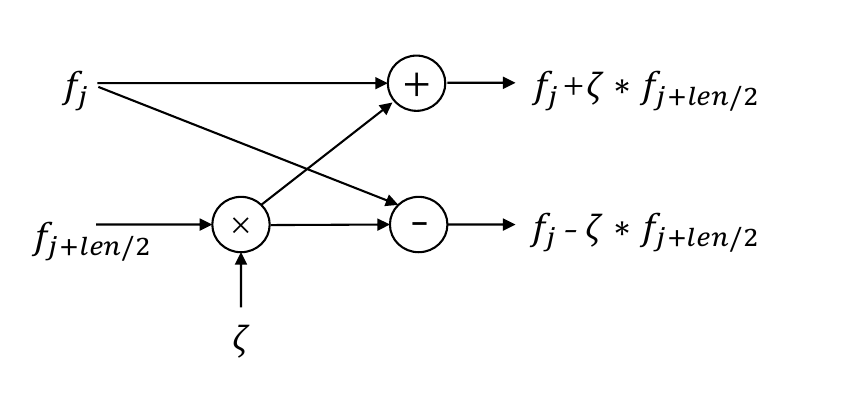}
\caption{Cooley-Tukey Butterfly.}
\label{fig:ct}
\end{minipage}
\begin{minipage}[t]{0.48\textwidth}
\centering
\includegraphics[width=6cm]{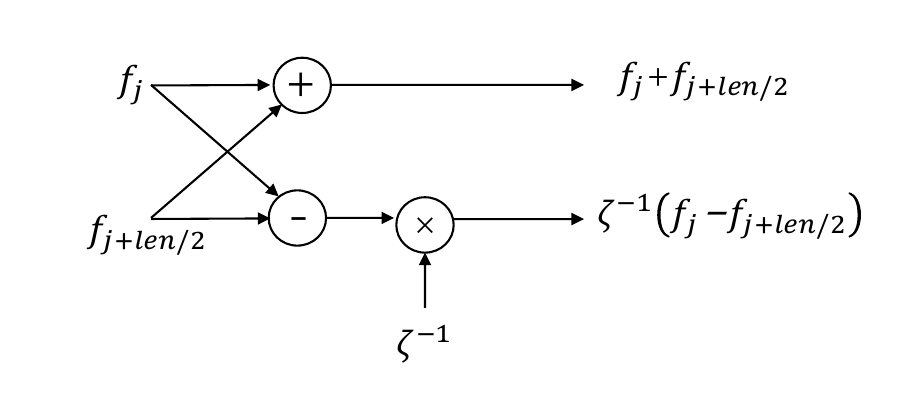}
\caption{Gentleman-Sande Butterfly.}
\label{fig:gs}
\end{minipage}
\end{figure}

Cooley-Tukey transform \cite{cooley1965algorithm} and Gentleman-Sande transform \cite{gentleman1966fast} are used in \textsf{NTT} and \textsf{INTT} respectively, shown in Figure \ref{fig:ct} and Figure \ref{fig:gs}.

\subsection{AVX-512 Instruction Set}
The AVX-512 instruction set was introduced by Intel in 2013 and initially supported in the 2016 Xeon Phi series processors \cite{intel-xeon-phi-processor}, AVX-512 has since become a pivotal feature for both major CPU manufacturers. AVX-512 offers several key functionalities: it introduces 32 512-bit \texttt{zmm} registers, enabling simultaneous processing of multiple data elements and accelerating vectorized computations. AVX-512 covers a wide range of floating-point and integer operations for diverse computational requirements. Additionally, AVX-512 provides 8 mask registers (\texttt{k0-k7}) for conditional operations, allowing instructions to execute based on conditions. These mask registers enhance the flexibility and efficiency of data processing by enabling advanced operations such as compression/expansion and masking. 

\subsection{TLS 1.3 and PQ-TLS}
TLS is a standard developed by the Internet Engineering Task Force (IETF) in 1999. Its primary role is to encrypt communication between web applications and servers, with the latest version being TLS 1.3 \cite{rescorla2018tls}. When initiating a TLS connection, the client and server exchange parameters such as TLS version and ciphersuite. Our study focuses on enhancing TLS security against post-quantum threats, particularly through the adoption of PQC Key Exchange (KEX) in PQ-TLS. PQ-TLS operates in two modes: hybrid mode and PQ-only mode. The hybrid mode, as standardized by the IETF in TLS 1.3 \cite{draft-ietf-tls-hybrid-design}, supports simultaneous usage of ECDH and PQC KEM, albeit at the cost of increased data transmission size and computational resources. Conversely, the PQ-only mode exclusively employs PQC KEM for Key Exchange and PQC signature algorithms for authentication.
\section{ML-KEM AVX-512 Implementation}
\label{sec:pqc-avx512}
The most time-consuming operations in \textsf{ML-KEM} include modular reduction, polynomial sampling, polynomial multiplication, and hash functions. In this section, we will outline our rationale behind the AVX-512 implementation design for these computationally intensive components.

\subsection{Modular Reduction Implementation}

Modular reduction plays a crucial role in \textsf{ML-KEM} polynomial arithmetic, where the arithmetic operates over the ring $\mathbb{Z}_q$ with $q=3329$. However, AVX-512 instructions lack dedicated support for modular reduction computations. In this context, we introduce the signed versions of two constant-time reduction algorithms commonly utilized in lattice-based cryptography: Montgomery reduction \cite{montgomery1985modular} and Barrett reduction \cite{barrett1986implementing}, as proposed by Seiler \cite{DBLP:journals/iacr/Seiler18}.\\
\textbf{Reduction Algorithms.}
The Signed Montgomery reduction computes the Hensel remainder of a signed integer within the range $[-\beta q/2,\beta q/2)$, specifically employed in \textsf{ML-KEM} to reduce the product of two 16-bit coefficients. The output remainder integer resides in the Montgomery domain and is further multiplied by $\beta^{-1}$. Barrett reduction operates within a smaller input range compared to Montgomery reduction. Typically, the input to Barrett reduction does not exceed a 16-bit signed integer range. Therefore, Barrett reduction is commonly utilized to reduce coefficients that surpass the range $[-q,q]$ after addition and subtraction operations. The resulting output from Barrett reduction remains within $\mathbb{Z}_q$.\\
\textbf{AVX-512 Implementations.}
We give our AVX-512 implementations of signed Montgomery reduction and Barrett reduction. 
For the modulus $q=3329=2^{13}-2^9+1$, we set $\beta=2^{16}$. This configuration ensures that one coefficient occupies 16 bits within a 512-bit vector register, enabling 32-way parallelism. Given that $v$ remains constant in Barrett reduction \cite[Sec 3.3]{DBLP:journals/iacr/Seiler18}, we can precompute $v$ and store it in the vector register. Regarding the computation of $av/2^\beta$, it involves two operations: multiplication and division. These two operations can be executed using a single AVX-512 instruction, \texttt{vpmulhw}, which computes the product of every two 16-bit data lanes in vector registers and retains only the higher 16 bits. Additionally, the division by $2^{\lfloor\log (q)\rfloor-1}$ can be efficiently implemented using AVX-512 shift instructions.  In Algorithm \ref{algo:barrettreductionavx512}, we introduce a macro \texttt{red16} designed to compute Barrett reduction using AVX-512 instructions. Within this macro, the \texttt{zmm\textbackslash r} register holds the output of the Barrett reduction, while \texttt{zmm\textbackslash rv} stores the constant used in Barrett reduction, represented by $v=\left\lfloor\frac{2^{\lfloor\log (3329)\rfloor-1} 2^{16}}{3329}\right\rceil=20159$. Additionally, the \texttt{zmm\textbackslash rl} register stores an immediate value. Furthermore, we will introduce our 32-way Montgomery reduction AVX-512 implementation, combined with butterfly operation, in Section \ref{sec:NTT-avx512}.
\input{Algorithms/barrettavx512}

\subsection{NTT Implementation}
\label{sec:NTT-avx512}
NTT is one of the most intricate components in the \textsf{ML-KEM} AVX-512 implementation. Both \textsf{NTT} and \textsf{INTT} operations in \textsf{ML-KEM} require 7 layers of butterfly operations to obtain the final result. In this section, we will introduce several techniques used in NTT implementation. \\
\textbf{Register Allocation.}
A 512-bit vector register \texttt{zmm} can accommodate a maximum of 32 16-bit integers, requiring only 8 vector registers to store all 256 polynomial coefficients. Therefore, we allocate 8 vector registers for storing coefficients, 2 for intermediate registers, and 2 for constant registers, leaving 20 vector registers unused. \\
\textbf{Butterfly Unit.} Our Cooley-Tukey butterfly pseudo-code is outlined in Algorithm \ref{algo:CTavx512}. Registers \texttt{zmm\textbackslash l} and \texttt{zmm\textbackslash r} respectively store coefficients $f_i$ and $f_{j+len/2}$, while \texttt{zmm0} holds the modulus $q$. We precompute $q^{-1} \bmod ^{\pm} \beta$ into twiddle factors to reduce one multiplication operation in Montgomery reduction. To obtain the low and high bits of the 16-bit integer multiplication product, we utilize vector integer multiplication instructions \texttt{vpmullw} and \texttt{vpmulhw}. These instructions eliminate the necessity of extending coefficients to 32 bits after multiplication. Our Gentleman-Sande Butterfly pseudo-code is presented in Algorithm \ref{algo:gsavx512}. Similar to the Cooley-Tukey butterfly, registers \texttt{zmm\textbackslash l} and \texttt{zmm\textbackslash r} store coefficients $f_i$ and $f_{j+len/2}$, respectively. Additionally, registers \texttt{zmm\textbackslash zl} and \texttt{zmm\textbackslash zr} hold precomputed $\zeta \cdot q^{-1}$ and $\zeta$ values, respectively. The order of twiddle factors in the \textsf{NTT} differs from that in the \textsf{INTT}, thus the same twiddle factors can be used. However, employing the same twiddle factors table necessitates additional permutation. In our implementation, we opt to utilize two distinct twiddle factors tables for the \textsf{NTT} and \textsf{INTT}.

\input{Algorithms/GSavx512}
\input{Algorithms/CTavx512}

In the implementation of \textsf{NTT} and \textsf{INTT}, we employ layer merging and coefficient permutation methods to reduce the memory access overhead. Specifically, we merge the 7 layers of NTT. We load all coefficients into 8 vector registers only at the first layer and store them back in memory after completing the final layer computation. Achieving such layer merging incurs some additional overhead for the coefficient permutation. Since each vector register accommodates 32 coefficients, in the initial three layers of the \textsf{NTT}, the coefficients stored in vector registers satisfy the correct distance and can directly perform the Cooley-Tukey butterfly. Starting from the fourth layer, as the butterfly distance becomes 16, the coefficient pairs requiring butterfly operations are housed within the same vector register. Thus, it's necessary to separate the coefficient pairs at corresponding distances into different vector registers. 
\begin{figure}[!htbp]
  \centering
  \includegraphics[width=1.0\linewidth]{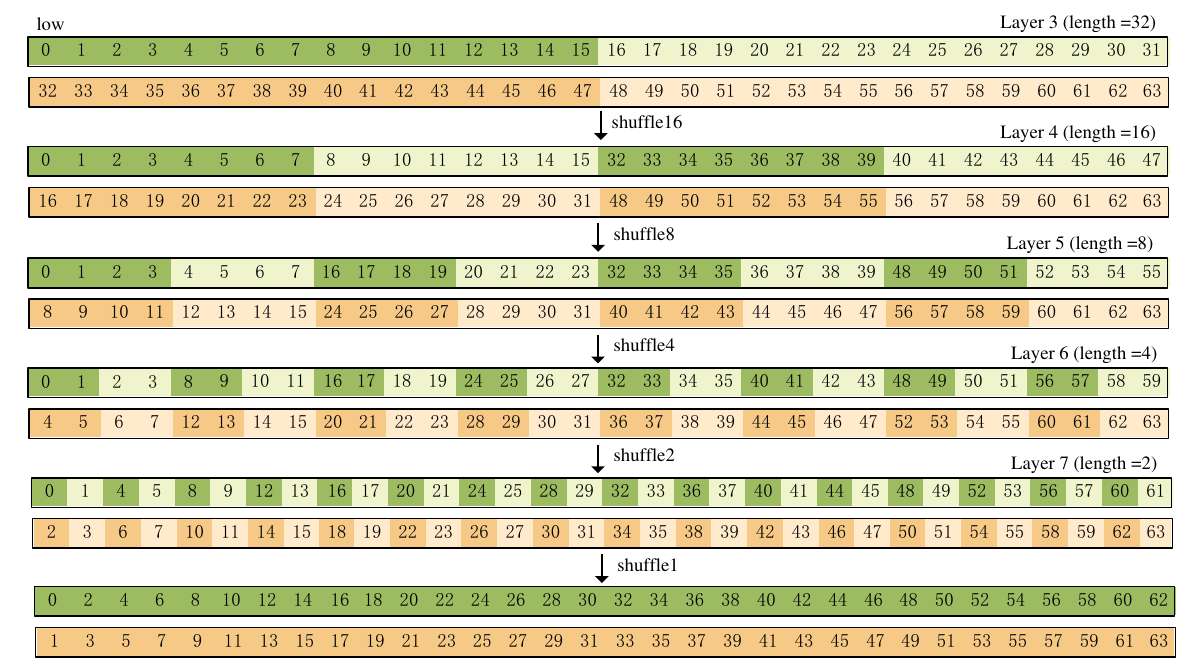}
  \caption{Coefficients permutation.}
  \label{fig:shuffle}
\end{figure}
As depicted in Figure \ref{fig:shuffle}, we can observe the arrangement of coefficients within a vector register during the third to seventh layers of the NTT. The functions \textsf{Shuffle16}, \textsf{Shuffle8}, \textsf{Shuffle4}, \textsf{Shuffle2}, and \textsf{Shuffle1} are utilized for permuting pairs of 16, 8, 4, 2, and 1 coefficients, respectively.  We perform an extra \textsf{Shuffle1} to bring convenience for polynomial point-wise multiplication.  After completing all \textsf{NTT} layers, the coefficients are constrained within the range of signed 16-bit integers, eliminating the need for Barrett reduction in \textsf{NTT}. However, this differs in the case of \textsf{INTT}. We adopt lazy reduction in \textsf{INTT} as outlined in \cite{westerbaan2020barrett}.


\subsection{SHA3 Keccak Implementation}
\label{sec:parkeccak}
\textsf{ML-KEM} utilizes \textsf{SHA3-256}, \textsf{SHA3-512}, \textsf{SHAKE128}, and \textsf{SHAKE256} as hash functions, pseudorandom functions (PRFs), and eXtendable-output functions (XOFs). These algorithms belong to the SHA-3 family \cite{fips2014secure} developed by NIST and are based on the Sponge Construction \cite{bertoni2007sponge}. The round function employed in the SHA-3 algorithm is the \textsf{Keccak-p}[1600,24] permutation function, which operates on a 64-bit state. In the previous AVX2 implementation of \textsf{ML-KEM} by the Crystals team, \textsf{Keccak-p}[1600,24] achieved 4-way parallelism due to the 256-bit vector register width. However, with AVX-512 implementation, we can compute 8 Keccak permutation results in parallel, as a 512-bit vector register can process 8 Keccak states concurrently. 
\subsection{Other Modules}
We also implement \textsf{Compress} and \textsf{Decompress}, \textsf{pkDecode}, \textsf{skDecode}, \textsf{skEncode} and \textsf{pkEncode} and several polynomial arithmetic functions. The implementation idea is similar to the AVX2 implementation. For simplicity, we don't discuss details here. One notable difference is that, in the implementation of \textsf{pkDecode}, \textsf{skDecode}, and \textsf{Decompress}, we utilize the AVX-512 instruction \texttt{vpermb}, which operates on bytes, unlike AVX2 instructions. This instruction allows us to conveniently adjust the order of byte streams. We implement rejection sampling using AVX-512 according to the method presented in \cite{zheng2023optimized}.

\section{ML-KEM TLS 1.3 Integration Design Consideration}
In this section, we discuss how to integrate \textsf{ML-KEM} AVX-512 implementation into TLS 1.3.

\subsection{Batch Key Generation Using Parallel Keccak}
\label{sec:batch}

The concept of batch key generation was discussed in both \cite{zhang2024eng25519} and \cite{DBLP:conf/uss/BernsteinBCT22}. In \cite{DBLP:conf/uss/BernsteinBCT22}, Montgomery's trick was used to compute multiple polynomial inversions, specifically targeting key generation algorithms involving polynomial inversion. On the other hand, \cite{zhang2024eng25519} introduces an $8\times1$ \textsf{X25519} approach to batch 8 key pairs in parallel. We explore the potential applicability of batch key generation in other PQC algorithms. Drawing inspiration from \cite{roy2019saberx4}, we propose a batch key generation approach for \textsf{ML-KEM}, as outlined in Algorithm \ref{algo:batchkemkeygen}. The \textsf{SHA3-256x8} function is constructed based on the 8-way Keccak function discussed in Section \ref{sec:parkeccak}. Our 8-way \textsf{ML-KEM} key generation AVX-512 implementation function is designed to generate 8 independent $(pk,sk)$ pairs simultaneously.

\input{Algorithms/batchml-kemgen}

\subsection{ML-KEM AVX-512 TLS 1.3 Migration Implementation}
We use OQS provider proposed by the Open Quantum Safe (OQS) team \cite{OQS} and OpenSSL 3.3.0-dev to migrate the \textsf{ML-KEM} AVX-512 implementation. OpenSSL is an open-source software library implementing the SSL and TLS protocols. Most PQ-TLS research works use OQS-OpenSSL to migrate PQC algorithms into TLS 1.3. However, OQS-OpenSSL ceased updates in July 2023 and does not support the latest OpenSSL 3.0 version. The latest OQS provider separates the integration of PQ algorithms into TLS 1.3 from the main logic of OpenSSL, without altering the core cryptographic algorithms. This separation isolates the embedding of post-quantum algorithms from the extensive OpenSSL codebase.  For \textsf{ML-KEM} AVX-512 code migration, we choose liboqs 0.10.1 \cite{LIBOQS}. Liboqs is an open-source C library for quantum-safe cryptographic algorithms. The newest liboqs 0.10.1 version adds \textsf{ML-DSA} and \textsf{ML-KEM} C reference and AVX2 codes. We add our \textsf{ML-KEM} AVX-512 code in the \texttt{ml\_kem} directory of liboqs. Besides, we implement the corresponding \textsf{ML-KEM} AVX-512 \textsf{Keygen}, \textsf{Encaps}, and \textsf{Decaps} APIs in liboqs. We define the macro \texttt{OQS\_ENABLE\_KEM\_ml\_kem\_512\_avx512} in the OQS configuration file. By configuring this macro, users can run \textsf{ML-KEM-512} AVX-512 code within the liboqs library.



\section{Revisiting PQC Security of KEM in TLS 1.3}

In this section, we revisit PQC security of KEM in TLS 1.3 based on recent research \cite{DBLP:conf/eurocrypt/Huguenin-Dumittan22,jiang2023post}. We state that the IND-CCA KEM used in TLS 1.3 handshake can be replaced by an IND-1-CCA KEM, providing improved efficiency and sufficient security.

\subsection{An Efficient Choice of Key-Exchange: IND-1-CCA KEM}

In existing PQ-TLS implementations, the ephemeral KEX is implemented with IND-CCA KEMs. IND-CCA KEMs are usually constructed by applying Fujisaki-Okamoto (FO) transform or its variants (e.g. \textsf{ML-KEM} \cite{NISTFIPS203}) on an OW/IND-CPA PKE scheme, while FO transform requires re-encrypting the plaintext during decapsulation, significantly reducing the efficiency of KEMs and increasing the cost of side-channel protection \cite{Melissa2022slides}.

Recent protocols (e.g. TLS 1.3 and KEMTLS) are designed to achieve forward security. In such protocols, each pair of ephemeral public/private keys is discarded immediately after being used once, and a new key pair will be generated for new messages. This means that an adversary will be able to request a decryption only once for a given key pair. Informally, IND-1-CCA security states that an adversary needs to distinguish an honestly generated key from a randomly generated key with at most one decapsulation query. Thus, the IND-1-CCA security of KEMs is sufficient to replace the Diffie-Hellman (DH) key-exchange, ensuring the security of such protocols. In the security proof of TLS 1.3 handshake under the multi-stage model given in \cite{dowling2021cryptographic}, the DH key-exchange could be replaced by an IND-1-CCA KEM and the proof would still hold. This idea inspired a series of work, see \cite{DBLP:conf/eurocrypt/Huguenin-Dumittan22,jiang2023post,schwabe2020post,schwabe2021more} for details.

\cite{DBLP:conf/eurocrypt/Huguenin-Dumittan22} proposed that an IND-$q$-CCA-secure KEM could be obtained from any passively secure PKE (OW-CPA/IND-CPA) without re-encryption. Specifically, \cite{DBLP:conf/eurocrypt/Huguenin-Dumittan22} presented two constructions named ${T_{CH}}$ and ${T_H}$, as well as their security proofs. Both ${T_{CH}}$ and ${T_H}$ do not require re-encryption and de-randomization, and such IND-1-CCA KEMs could be used in TLS 1.3 handshake, improving the efficiency while ensuring security. However, ${T_{CH}}$ leads to ciphertext expansion, and the security of ${T_H}$ was only proved in ROM, the QROM proof was not provided. Based on \cite{DBLP:conf/eurocrypt/Huguenin-Dumittan22}, \cite{jiang2023post} provided both ROM and QROM proofs of ${T_H}$ and ${T_{RH}}$ (an implicit variant of ${T_H}$), with much tighter reductions than \cite{DBLP:conf/eurocrypt/Huguenin-Dumittan22}.

\subsection{IND-1-CCA KEM Constructions}

\begin{figure}[htbp]
  \centering
  \includegraphics[width=1.0\linewidth]{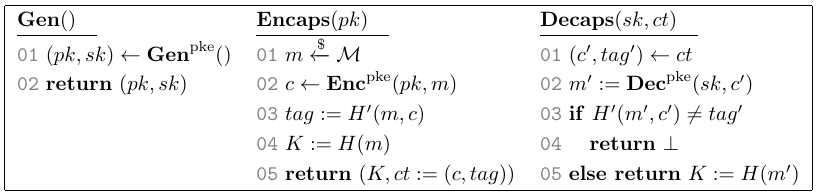}
  \caption{$T_{CH}$ transform.}
  \label{fig:TCH}
\end{figure}

\begin{figure}[htbp]
  \centering
  \includegraphics[width=1.0\linewidth]{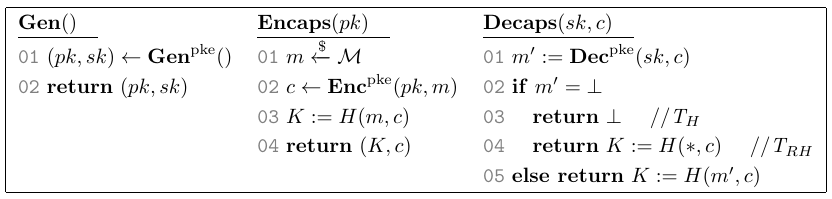}
  \caption{$T_{H}/T_{RH}$ transform.}
  \label{fig:TH}
\end{figure}

In this section, we review the definition of ${T_{CH}}$ and ${T_H}/{T_{RH}}$, as well as their reduction tightness in ROM/QROM.\\
\textbf{Constructions. }The idea of ${T_{CH}}$ is to send an additional hash value along with ciphertext, which will be used for key confirmation in decapsulation (see Figure \ref{fig:TCH}). ${T_H}$ works without additional hash value, the key is derived by $H(m,c)$ directly (Figure \ref{fig:TH}). ${T_{RH}}$ is an implicit variant of ${T_H}$, the only difference between ${T_{RH}}$ and ${T_H}$ is the return value for invalid decryption. In fact, ${T_H}$ and ${T_{RH}}$ is equivalent to ${U^ \bot }$ and ${U^ {\not\bot}}$ in \cite{hofheinz2017modular}, respectively.\\
\textbf{Reduction Tightness. }The security of ${T_{CH}}$ was proved in ROM and QROM by Huguenin-Dumittan and Vaudenay \cite{DBLP:conf/eurocrypt/Huguenin-Dumittan22}, with tightness ${\epsilon _R} \approx O(1/{q_H}){\epsilon _{\cal A}}$ and ${\epsilon _R} \approx O(1/{{q_H}^3})\epsilon _{\cal A}^2$ respectively, where ${\epsilon _R}$ (resp. ${\epsilon _{\cal A}}$) is the advantage of the reduction $R$ (resp. adversary ${\cal A}$) against the underlying PKE (resp. the IND-1-CCA KEM) and ${q_H}$ denotes the number of ${\cal A}$'s queries to $H$. Later, they updated an ePrint version of \cite{DBLP:conf/eurocrypt/Huguenin-Dumittan22} and gave a tighter QROM proof of ${T_{CH}}$ with bound ${\epsilon _R} \approx O(1/{{q_H}^2})\epsilon _{\cal A}^2 - O({{q_H}^3}/{2^n}) - O({q_H}/\sqrt {{2^n}} )$. The ROM and QROM proof of ${T_H}$ and ${T_{RH}}$ were presented in \cite{jiang2023post}, with tightness ${\epsilon _R} \approx O(1/{q_H}){\epsilon _{\cal A}}$ and ${\epsilon _R} \approx O(1/{{q_H}^2})\epsilon _{\cal A}^2$ respectively. The comparison between FO transform and ${T_{CH}}/{T_{H}}/{T_{RH}}$ is summarized in Table \ref{tab:FOCompar}.
 \\
\input{Tables/FO-likeComparison}


\section{Discussions and Results}
\subsection{Experimental Setup}
In this section, we describe our experimental setup. We used an emulation experiment setup in one machine. The benchmark platform is an Intel Xeon Platinum 8475B with 4 vCPUs and 8 GiB memory \footnote{Our testing machine is provided by Alibaba Cloud, belonging to the compute-optimized instance family c8i. It is powered by an Intel Xeon Sapphire Rapids processor, with the specific specification ecs.c8i.xlarge, featuring 4 vCPUs and 8 GiB of memory.}. The OpenSSL version we used is OpenSSL3.3.0-dev. We used GCC 9.4.0 to compile all programs. We disabled AVX-512 instructions when benchmarking ML-KEM AVX2 implementation for better comparison. We didn't disable Turbo Boost and hyper-threading to simulate real-world conditions during TLS handshake.

\subsection{Speed of ML-KEM AVX-512 Implementation}
In this section, we conduct a performance evaluation of our \textsf{ML-KEM} AVX-512 implementations. Our approach is built upon the Crystals team's open-source Kyber Standard Code \cite{kyberstandard}, which was implemented for the FIPS 203 draft. To assess performance, we executed the speed testing program 100,000 times and recorded the median CPU cycle count. The results are summarized in Table \ref{tab:avx512performance_comparison}, which compares various \textsf{ML-KEM} implementations across three security levels. Our implementation achieved a speedup ratio ranging from approximately 1.30$\times$ to 1.64$\times$ compared to the state-of-the-art AVX2 implementation. This performance enhancement can be attributed to the intricate design of our NTT AVX-512 implementation and the vectorization benefits offered by AVX-512 instructions. As illustrated in Table \ref{tab:avx512performance_comparison}, adopting batch method in Section \ref{sec:batch} can accelerate the key generation process by 3.5$\times$ to 4.9$\times$.

\input{Tables/AVX512benchmark}

\subsection{ML-KEM AVX-512 TLS 1.3 Benchmark}
We evaluated the performance of TLS 1.3 handshakes integrated with the \textsf{ML-KEM} AVX-512 implementation in a simulated network environment. Initially, we utilized OpenSSL commands to generate key pairs and server certificates. Subsequently, utilizing the previously generated keys and certificates, we launched a TLS server via the \texttt{openssl s\_server} tool \cite{opensslsserver}. We assessed the handshake connections per second using the \texttt{openssl s\_time} tool \cite{opensslstime}. Both the server and client were compiled against OpenSSL3.3.0-dev with the OQS provider, thus facilitating support for both hybrid mode and PQ-only mode.\\
\textbf{PQ-only ML-KEM PQ-TLS 1.3 Benchmark. }
In TLS, we mainly observe the \textsf{ML-KEM} algorithm for KEX. However, there are multiple signature algorithms available, each with different security levels. Therefore, observing the combination of different signature algorithms with \textsf{ML-KEM} is crucial and worth considering. For our experiments, we selected three signature algorithms: the standardized post-quantum signature algorithms \textsf{Dilithium} and \textsf{Falcon}, as well as the traditional public key cipher \textsf{RSA2048}.

\input{Tables/handshake}

As shown in Table \ref{tab:pqhandshake}, under PQ-only setting, AVX-512 implementation of \textsf{ML-KEM} brings more handshakes per second compared to AVX2 implementation. This indicates that using AVX-512 to optimize \textsf{ML-KEM} performance can bring improvement to the PQ-TLS handshake. However, the improvement is subtle. Since handshake time is not solely determined by KEX time but also influenced by factors such as transmission time and authentication time, improving the performance of the \textsf{ML-KEM} algorithm does not lead to a significant enhancement in overall handshake time performance. Besides, the handshake time decreases as the security level increases. This is in line with expectations because higher security levels typically result in a slower speed of signature and KEM. Overall, using \textsf{Dilithium} as the signature authentication scheme performs better in handshakes compared to \textsf{Falcon} under the same security level.  Compared to \textsf{RSA2048}, using the \textsf{Dilithium} and \textsf{Falcon} post-quantum signature schemes for authentication does not impose significant performance overhead. These conclusions provide insights into the impact of different signature algorithms combined with \textsf{ML-KEM} on TLS handshake performance and serve as a reference for selecting the optimal solution. \\
\textbf{Hybrid ML-KEM PQ-TLS 1.3 Benchmark. } 
OQS provider facilitates hybrid key exchange for TLS 1.3 by the IETF draft on Post-Quantum Traditional Hybrid Schemes \cite{ietf-hybrid-terminology}. Consequently, we also evaluated the performance of \textsf{ML-KEM} and other ECDH curves hybrid schemes. The data from Table \ref{tab:hybridhandshake} indicates a significant decrease in handshake performance in hybrid mode compared to the PQ-only mode. Furthermore, the enhancement in handshake performance achieved through the AVX-512 implementation is minimal. This phenomenon may be attributed to the fact that in hybrid mode, the handshake time is primarily composed of two components: \textsf{ECDH} and \textsf{ML-KEM}. Therefore, solely improving the performance of \textsf{ML-KEM} in hybrid mode is insufficient; there is also a need to enhance the performance of the \textsf{ECDH} scheme.
\input{Tables/hybridhandshake}

\subsection{TLS 1.3 Handshake with IND-1-CCA KEM}
In this section, we implemented IND-1-CCA KEMs based on ${T_{CH}}$ and ${T_{RH}}$ constructions proposed in \cite{DBLP:conf/eurocrypt/Huguenin-Dumittan22} and \cite{jiang2023post}, integrated the better one into TLS 1.3 and analyzing its performance.\\
\textbf{Comparison of KEM Constructions.}
We constructed two IND-1-CCA KEMs based on the underlying PKE of ML-KEM, employing ${T_{CH}}$ and ${T_{RH}}$ respectively. Both ${T_{CH}}$ and ${T_{RH}}$ utilized \textsf{SHA3-256} to instantiate the hash function \textsf{H}, consistent with the original $FO$ transform in \textsf{ML-KEM}. Additionally, we fixed the length of the $tag$ in ${T_{CH}}$ to 32 bytes.  The results are summarized in Table \ref{tab:fobenchmark}, which compare the execution speed and communication overhead across three constructions (\textsf{ML-KEM}'s original $FO$ construction, ${T_{CH}}$ and ${T_{RH}}$). From the experiment results, the following facts are concluded.
 For key-generation and encapsulation, ${T_{CH}}$ and ${T_{RH}}$  show rational improvement in efficiency, mainly because they omit the hash of $pk$ in key-generation, as well as the de-randomization in encapsulation.  For Decapsulation, ${T_{CH}}$ and ${T_{RH}}$ achieve a speed ratio of at least 3.04$\times$ compared to $FO$ implementation, which can be attributed to removing re-encryption. The improvement becomes more significant as the security level increases because higher security levels result in slower encryption.  ${T_{CH}}$ requires extra 32 bytes in ciphertext storage, and ${T_{CH}}$'s decapsulation is slightly slower than ${T_{RH}}$ due to the key confirmation.  Overall, KEMs based on ${T_{RH}}$ show better performance in both efficiency and communication overhead.
 
\input{Tables/fobenchmark}
\noindent \textbf{IND-1-CCA KEM TLS 1.3 Benchmark.}
The ${T_{RH}}$-based IND-1-CCA KEM is integrated into TLS 1.3 in both PQ-only mode and hybrid mode. We measured the number of handshakes per second and compared the results with the original $FO$ implementation, shown in Table \ref{tab:fohandshake}. 
\input{Tables/FOhandshake}

The results show that ${T_{RH}}$-based IND-1-CCA KEMs increase the number of TLS 1.3 handshakes per second while maintaining communication overhead. Especially in the PQ-only mode with a higher security level, the removal of re-encryption brings more significant improvement, which is in line with our expectations.

Our experiments confirm the advantages of applying IND-1-CCA KEMs in TLS 1.3 at a practical level. Specifically, in application scenarios that place more emphasis on handshake efficiency, IND-1-CCA KEMs have better adaptability than $FO$-based IND-CCA KEMs. These results inspire further research on IND-1-CCA constructions (e.g. tighter reduction) and their TLS 1.3 implementation.
\section{Conclusion}
In this paper, we present our implementation of \textsf{ML-KEM} using AVX-512 and introduce a novel batch method for \textsf{ML-KEM} key generation. With the support of OQS provider, we seamlessly integrate our optimized \textsf{ML-KEM} AVX-512 implementation into TLS 1.3, enhancing its resistance to post-quantum threats. We evaluate the performance of our implementation in both PQ-only and hybrid modes. Furthermore, we revisit two IND-1-CCA KEMs and analyze their impact on PQ-TLS handshake performance. Our AVX-512 implementation demonstrates a speedup of up to 1.64$\times$ for \textsf{ML-KEM}, while our batch method achieves significant speedups ranging from 3.5$\times$ to 4.9$\times$ for key generation. Furthermore, our measurements show improvements in TLS 1.3 handshake performance with our AVX-512 implementation. Through our optimized implementation, integration, and assessment efforts, we provide valuable insights for future work aimed at enhancing PQ-TLS handshake performance.
\bibliographystyle{splncs04}
\bibliography{ref}
\end{sloppypar}
\end{document}

%% file: Algorithms/barrettavx512.tex
\begin{algorithm}[htbp]
  \caption{Signed Barrett reduction AVX-512 code}
  \label{algo:barrettreductionavx512}
  \small
  \begin{algorithmic}[1]
    \State \texttt{.macro   red16   r, rv, rl}
    \State \texttt{vpmulhw     \%zmm\textbackslash rv, \%zmm\textbackslash r, \%zmm\textbackslash rl}
    \State \texttt{vpsraw       \$10, \%zmm\textbackslash rl, \%zmm\textbackslash rl}
    \State \texttt{vpsubw      \%zmm\textbackslash rl, \%zmm\textbackslash r, \%zmm\textbackslash r}
    \State \texttt{.endm}
  \end{algorithmic}
\end{algorithm}

%% file: Algorithms/GSavx512.tex
\begin{algorithm}[!htbp]
   
  \caption{Gentleman-Sande butterfly AVX-512 code}
  \label{algo:gsavx512}
  \small
  \begin{algorithmic}[1]
    \State \texttt{.macro   Gentleman-Sande butterfly   l,r,zl,zh}
    \State \texttt{vpsubw      \%zmm\textbackslash l,\%zmm\textbackslash r,\%zmm21}
    \State \texttt{vpaddw      \%zmm\textbackslash r,\%zmm\textbackslash l,\%zmm\textbackslash l}
    \State \texttt{vpmulhw     \%zmm\textbackslash  zh,\%zmm21,\%zmm22}
    \State \texttt{vpmulhw     \%zmm\textbackslash zh,\%zmm21,\%zmm22}

     \State \texttt{vpmulhw \%zmm0,\%zmm21,\%zmm21}
     \State \texttt{vpsubw \%zmm21,\%zmm22,\%zmm\textbackslash r}
    \State \texttt{.endm}
  \end{algorithmic}
\end{algorithm}

%% file: Algorithms/CTavx512.tex
\begin{algorithm}[htbp]
  \caption{Cooley-Tukey butterfly AVX-512 code}
  \label{algo:CTavx512}
  \small
  \begin{algorithmic}[1]
    \State \texttt{.macro  Cooley-Tukey butterfly   l,r,zl,zh}
    \State \texttt{vpmullw		\%zmm\textbackslash zl,\%zmm\textbackslash r,\%zmm11}
    \State \texttt{vpmulhw		\%zmm\textbackslash zh,\%zmm\textbackslash r,\%zmm\textbackslash r}
     \State \texttt{vpmulhw		\%zmm0,\%zmm11,\%zmm11}
     \State \texttt{vpsubw      \%zmm11,\%zmm\textbackslash l,\%zmm\textbackslash r}
     \State \texttt{vpaddw      \%zmm11,\%zmm\textbackslash l,\%zmm\textbackslash l}
    \State \texttt{.endm}
  \end{algorithmic}
\end{algorithm}

%% file: Algorithms/batchml-kemgen.tex
\begin{algorithm}[htbp]
  \caption{Batch key generation for \textsf{ML-KEM.KEM}}
  \label{algo:batchkemkeygen}
  \small
  \begin{algorithmic}[1]
    \Ensure Encapsulation keys array {ek[8]} $\in\mathbb{B}^{384k+32}$.
    \Ensure Decapsulation keys array {dk[8]} $\in\mathbb{B}^{768k+96}$.
    \State $z \stackrel{\$}{\longleftarrow} \mathbb{B}^{32}$
     \For {i = 0 to 7}
    \State $\left(\mathrm{e} \mathrm{k}_{\mathrm{PKE}}[i], \mathrm{d} \mathrm{k}_{\mathrm{PKE}}[i]\right) \leftarrow$ K-PKE.KeyGen()
    \State $\mathrm{ek}[i] \leftarrow \mathrm{e} \mathrm{k}_{\mathrm{PKE}}[i]$
     \EndFor
    \State output array $output[8]$ $\leftarrow$ \textsf{SHA3-256x8}$(\mathrm{ek}[0],\mathrm{ek}[1],\mathrm{ek}[2],\mathrm{ek}[3],\mathrm{ek}[4],\mathrm{ek}[5],\mathrm{ek}[6],\mathrm{ek}[7])$
     \For {i = 0 to 7}
    \State $(\mathrm{dk}[i]\leftarrow\left(\mathrm{dk}_{\mathrm{PKE}}\|\mathrm{ek}\| output[i]) \| z\right)$
     \EndFor
    
    \State \textbf{Return} ({ek[8]}, {dk[8]})
  \end{algorithmic}
\end{algorithm}

%% file: Tables/FO-likeComparison.tex
\begin{table}
	\centering
 \renewcommand{\arraystretch}{1.5}  
	\fontsize{7.5}{12}\selectfont    
	\caption{Comparison between FO transform and ${T_{CH}}/{T_{H}}/{T_{RH}}$.}
    \resizebox{\linewidth}{!}{
         \begin{tabular}{m{1.6cm}<{\centering} m{3.1cm}<{\centering} m{2cm}<{\centering} m{1cm}<{\centering} m{2.5cm}<{\centering} m{2.5cm}<{\centering}}
		\toprule
	  Construction  & Security & $c$-Expansion & Re-Enc & \makecell{ROM \\ tightness} & \makecell{QROM \\ tightness} \\ 
		\hline
		$FO$ & IND-CPA $\Rightarrow$ IND-CCA & $\Circle$ & $\CIRCLE$ & ${\epsilon _R} \approx {\epsilon _{\cal A}}$ & ${\epsilon _R} \approx O(1/{q_H})\epsilon _{\cal A}^2$\\
		${T_{CH}}$ & OW-CPA $\Rightarrow$ IND-1-CCA & $\CIRCLE$ & $\Circle$ & ${\epsilon _R} \approx O(1/{q_H}){\epsilon _{\cal A}}$ & ${\epsilon _R} \approx O(1/{{q_H}^2})\epsilon _{\cal A}^2$ \\
		${T_{H}}/{T_{RH}}$ & IND-CPA $\Rightarrow$ IND-1-CCA & $\Circle$ & $\Circle$ & ${\epsilon _R} \approx O(1/{q_H}){\epsilon _{\cal A}}$ & ${\epsilon _R} \approx O(1/{{q_H}^2})\epsilon _{\cal A}^2$   \\
		\bottomrule
	\end{tabular}\vspace{0cm}
    }
       
	\label{tab:FOCompar}
\end{table}

%% file: Tables/AVX512benchmark.tex
\renewcommand{\arraystretch}{1} 
\begin{table}[htbp]  
	
	\centering 
 \renewcommand{\arraystretch}{1.5}  
	\fontsize{8}{8}\selectfont
  \resizebox{\linewidth}{!}{
	\begin{threeparttable}  
		\caption{ML-KEM implementation performance comparison. (CPU Cycles)}  
		\label{tab:avx512performance_comparison}  
  
		\begin{tabular}{ccccccccccc}  
			\toprule  
			\multirow{2}{*}{ }&  
			\multicolumn{3}{c}{ML-KEM-512}&\multicolumn{3}{c}{ ML-KEM-768}&\multicolumn{3}{c}{ ML-KEM-1024}&ISA/ISE\cr  
			\cmidrule(lr){2-4} \cmidrule(lr){5-7}   \cmidrule(lr){8-10}  
			&\textsf{Keygen}&\textsf{Encaps}&\textsf{Decaps}&\textsf{Keygen}&\textsf{Encaps}&\textsf{Decaps}&\textsf{Keygen}&\textsf{Encaps}&\textsf{Decaps}\cr  
			\midrule  
			 \cite{kyberstandard}&81750 
&90,922 
&118,790 
&130,048
&142,688
&178,206
&217,174
&223,780
&274,588&{x86-64}\cr  Speed-up
&6.21$\times$ 
&6.68$\times$ 
&7.80$\times$ 
&6.61$\times$ 
&7.61$\times$ 
&8.36$\times$ 
&8.61$\times$ 
&8.93$\times$ 
&9.52$\times$ 
\cr  
			\cite{kyberstandard}&17,750

&18,708
&19,838
&28,996
&28,566
&30,476
&39,996
&41,162
&44,538
&AVX2\cr  
Speed-up&1.35x

&1.37$\times$ 
&1.30$\times$ 
&1.47$\times$ 
&1.52$\times$ 
&1.43$\times$ 
&1.59$\times$ 
&1.64$\times$ 
&1.54$\times$ 
\cr  

			This work &13,170
&13,620
&15,230
&19,684
&18,760
&21,308
&25,216
&25,062
&28,834
&AVX-512\cr  
			Batch Keygen &2,804 (4.9$\times$)
&13,620
&15,230
&4,866 (4.2$\times$)
&18,760
&21,308
&7,394 (3.5$\times$)
&25,062
&28,834
&\cr 
	
			\bottomrule  
		\end{tabular}  

	\end{threeparttable}  
   }
\end{table}  

%% file: Tables/handshake.tex
\begin{table}[htbp]
\centering
\setlength{\tabcolsep}{5pt}       
\caption{PQ-only ML-KEM Handshake performance comparison.  (connections/sec)}
\label{tab:pqhandshake}
\fontsize{8}{12}\selectfont         
\begin{tabular}{l|c|c|c|c|c}
\hline
\multicolumn{1}{c|}{\multirow{2}{*}{\textbf{KEM/SIG}}} & \multicolumn{2}{c|}{\textbf{ML-KEM-512}} &\multicolumn{2}{c|}{\textbf{ML-KEM-768}} & \textbf{ML-KEM-1024}\\
\cline{2-6}
 &  \textbf{Dilithium2}&\textbf{Falcon512} & \textbf{RSA2048}&\textbf{Dilithium3} & \textbf{Dilithium5}\\
\hline
\textbf{AVX2} &3416.02 & 3191.46& 2582.38 & 2710.85& 3191.46\\
\hline
\textbf{AVX-512} & 3589.22 & 3364.61& 2650.45 & 2856.71&  2451.08\\
\hline
\end{tabular}
\end{table}


%% file: Tables/hybridhandshake.tex
\begin{table}[htbp]
\centering
\setlength{\tabcolsep}{8pt}       
\caption{Hybrid ML-KEM handshake performance using \textsf{RSA2048} for authentication. (connections/sec)}
\label{tab:hybridhandshake}
\fontsize{8}{12}\selectfont         
\begin{tabular}{l|c|c|c|c|c|c}
\hline
\multicolumn{1}{c|}{\multirow{2}{*}{\textbf{Hybrid KEM}}} & \multicolumn{2}{c|}{\textbf{ML-KEM-512}} & \multicolumn{2}{c|}{\textbf{ML-KEM-768}}  & \multicolumn{2}{c}{\textbf{ML-KEM-1024}}\\
\cline{2-7}
 \multicolumn{1}{c|}{} &  \textbf{p256}&\textbf{x25519} & \textbf{p384}&\textbf{x448} & \textbf{p521}& \textbf{p384}\\
\hline
\textbf{AVX2} & 893.98 & 1965.58& 375.19 & 948.04& 200.45& 369.84\\
\hline
\textbf{AVX-512} & 944.23 & 2031.37& 377.72 & 999.85&  201.66&  371.39\\
\hline
\end{tabular}
\end{table}

%% file: Tables/fobenchmark.tex
\begin{table}[htbp]
\centering
\fontsize{7}{12}\selectfont  
\setlength\tabcolsep{2.8pt} 
\caption{Performance comparison of different FO transform. (CPU cycles)}
\label{tab:fobenchmark}
\begin{tabular}{@{}lccccccccccccc@{}}
\toprule
& \multicolumn{4}{c}{\textbf{ML-KEM-512}} & \multicolumn{4}{c}{\textbf{ML-KEM-768}} & \multicolumn{4}{c}{\textbf{ML-KEM-1024}} \\
\cmidrule(lr){2-5} \cmidrule(lr){6-9} \cmidrule(lr){10-13}
& \textsf{Keygen} & \textsf{Encaps} & \textsf{Decaps} & \textsf{Ct} &\textsf{Keygen} & \textsf{Encaps} & \textsf{Decaps}& \textsf{Ct} & \textsf{Keygen} & \textsf{Encaps} & \textsf{Decaps} & \textsf{Ct} \\
\midrule
$FO$ & 17,750 & 18,708 & 19,838 & 768 & 28,996 & 28,566 & 30,476 & 1,088 & 39,996 & 41,162 & 44,538 & 1,568 \\
${T_{CH}}$ & 12,932 & 18,434 & 6,520 & 800 & 21,942 & 28,316 & 8,974 & 1,120 & 30,292 & 42,450 & 12,172 & 1,600 \\
${T_{RH}}$ & 12,896 & 17,676 & 5,666 & 768 & 21,942 & 27,580 & 8,440 & 1,088 & 30,336 & 40,096 & 11,240 & 1,568 \\
\bottomrule
\end{tabular}\\[5pt] 
\hspace*{-225pt}\textit{Note:} Ct represents Ciphertext.
\end{table}

%% file: Tables/FOhandshake.tex

\begin{table}[htbp]
 \setlength{\tabcolsep}{15pt}
\centering
\fontsize{7.5}{12}\selectfont 
\caption{ML-KEM handshake performance under different security. (connections/sec)}
\label{tab:fohandshake}
\begin{tabular}{ m{2cm} l c c} 
\toprule
    \textbf{SIG}       &         \textbf{KEM}           & $FO$      & ${T_{RH}}$     \\ \midrule
     & ML-KEM-512    & 3416.02  & 3476.17  \\
 Dilithium2  & p256\_ML-KEM-512    & 924.61  & 978.12  \\
         & x25519\_ML-KEM-512 & 2014.98 & 2038.19 \\
     \hline
         & ML-KEM-768         & 2710.85 & 2803.24 \\
      Dilithium3    &x25519\_ML-KEM-768         & 1942.31
 & 2036.58
 \\
         &p256\_ML-KEM-768        & 902.83
 & 931.99
 \\
    \bottomrule
\end{tabular}
\end{table}